\documentstyle[seceq,epsf,wrapft]{ptptex}

\notypesetlogo  
\preprintnumber[3cm]{
YITP98-20 \\ March 1998}

\title{
\Large \bf Angular Diameter Distances \\ in Clumpy Friedmann Universes}

\author{%
Kenji {\sc Tomita}
}

\inst{%
Yukawa Institute for Theoretical Physics \break 
Kyoto University, Kyoto 606-8502, Japan
}

\recdate{%
April 6, 1998
}

\abst{%
Solving null-geodesic equations, behavior of angular diameter distances is
studied in inhomogeneous cosmological models, which are given by
performing $N$-body simulations with the CDM spectrum. 
The distances depend on the separation angle $\theta$ of ray pairs,
the mass $m$ and the radius $r_s$ of particles
consisting of galaxies and dark matter balls, and cosmological model
parameters. The calculated distances are compared with the Dyer-Roeder 
angular diameter distance with the clumpiness parameter $\alpha \ (= 0 - 1)$,
and for each ray pair the corresponding $\alpha$ is
determined. Shooting many ray pairs, we derive statistical
quantities such as the average value ($\bar{\alpha}$), the dispersion
($\sigma_\alpha$), and the distribution of $\alpha$.   
 
It is found in four cosmological models with $(\Omega_0, \lambda_0) = 
(1, 0), (0.2, 0), (0.4, 0), (0.2,0.8)$ for the particle parameters $m 
\simeq 2 \times 10^{11}$ M$_{\odot}$ and $r_s = (10 - 40)h^{-1}$ kpc \ 
that (1) $\bar{\alpha}$ is nearly equal to 1 or the Friedmann distance is 
best fitted, \ (2) $\sigma_\alpha$ decreases with the increase of 
the redshift $z$ and radius $r_s$, \ (3) for $\theta >> 1$ arcsec,
$\sigma_\alpha$ is very small, but for $\theta < 1$ arcsec it is so large 
that the use of only the the Friedmann distance
($\alpha = 1$) may cause some errors for quantitative analysis of 
cosmological lensing, \ (4) the distribution of $\alpha$ is
symmetric and asymmetric for $\sigma_\alpha < 0.5$ and $> 0.8$,
respectively, and \ (5) $\sigma_\alpha$ is smallest and largest in
models with (1, 0) and (0.2, 0), respectively. The cosmological constant
has the role of decreasing $\sigma_\alpha$. 
}

\begin{document}

\maketitle

\section{Introduction}
Recently many kinds of lens phenomena are observed in strong and weak
forms of deformed images of quasars and galaxies due to lenses of
galaxies, clusters of galaxies, and large-scale matter distribution.
Lensing is being used as a new observational tool for remote dark
objects.

 For the analysis of such lens phenomena, angular diameter distances are
often used because of their convenience, but their definition is not 
unique. The most
representative ones are the Friedmann distance and Dyer-Roeder
distance, which are given as the distances in the homogeneous and
smooth matter distribution and the distance in low-density regions in
the inhomogeneous matter distribution, respectively.\cite{rf:dr1,rf:dr2}
The behavior of these distances has been discussed in
various papers.\cite{rf:kft,rf:wt,rf:as,rf:sef} It is important to 
clarify which angular diameter distance is most appropriate in
the observation of our Universe. This problem depends on
the separation angles $\theta$ of ray pairs, the mass $m$ and radius $r_s$
of lens objects, like galaxies and dark matter balls, and cosmological
model parameters $(\Omega_0, \lambda_0)$. 

In this paper we first use for this purpose the inhomogeneous models 
which were produced in previous papers,\cite{rf:t1,rf:t2}
using $N$-body simulations, and we then
consider many ray pairs which are made by solving null-geodesic
equations in the above mentioned models and from which the average 
properties of the angular diameter distances are derived, taking into 
account the values of $\theta, m, r_s, \Omega_0$ and $\lambda_0$. 
In \S 2 the behavior of
the Dyer-Roeder angular diameter distance with the clumpiness parameter
$\alpha$ in various background models is given explicitly. In \S 3 the
angular diameter distances in inhomogeneous models produced by the 
numerical simulations are derived. By comparing those with the 
Dyer-Roeder angular diameter distance, the value of $\alpha$ for each 
ray pair is determined, and its statistical behavior is presented. 
In \S 4 concluding remarks are given.  

\bigskip
\section{Dyer-Roeder angular diameter distance \\ with the clumpiness parameter
$\alpha$}

The Dyer-Roeder angular diameter distance $D_{\rm A}(z)$ satisfies
\begin{equation}
  \label{eq:ba1}
{d^2 \over dv^2} D_{\rm A} + {3 \over 2}(1+z)^5 \alpha \Omega_0 D_{\rm A} = 0,
\end{equation}
where $\alpha$ is the clumpiness parameter and $v$ is the
affine parameter.\cite{rf:kft,rf:wt,rf:as,rf:sef}  $v$ is related to $z$
by
\begin{equation}
  \label{eq:ba2}
dz/dv = (1 + z)^2 [(\Omega_0 z +1)(1+z)^2 - \lambda_0 z (2+z)]^{1/2}.
\end{equation}
The cases $\alpha \ = 0$ and $1$ represent the limiting
distances in the empty region and the Friedmann distance in the
homogeneous region.

As we consider the rays received by an observer at the present epoch,
we have the conditions
\begin{equation}
  \label{eq:ba3}
D_{\rm A}(0) = 0, \qquad (dD_{\rm A} /dz)_{z = 0} = c/H_0,
\end{equation}
where $H_0 \ (= 100 h$ km s$^{-1}$ Mpc$^{-1})$ is the Hubble 
constant.

Here we consider four models with  $(\Omega_0, \lambda_0) = 
(1, 0), (0.1, 0), (0.2, 0)$ and $(0.2,0.8)$. The above equations are
easily solved in each model and we obtain the $z$-dependence of $D_{\rm A}(z)$ 
for  $\alpha = 0$ and $1$ and the $\alpha$-dependence of
$D_{\rm A}(z)$ for $z = 0.5, 1, \cdots, 5$, which are shown in Figs. 1, 2 
and Figs. 3 $\sim$ 6, respectively. These dependences are used to interpret 
the results in the next section.

\begin{figure}[htb]
 \parbox{\halftext}{
\epsfxsize=7.5cm
\epsfbox{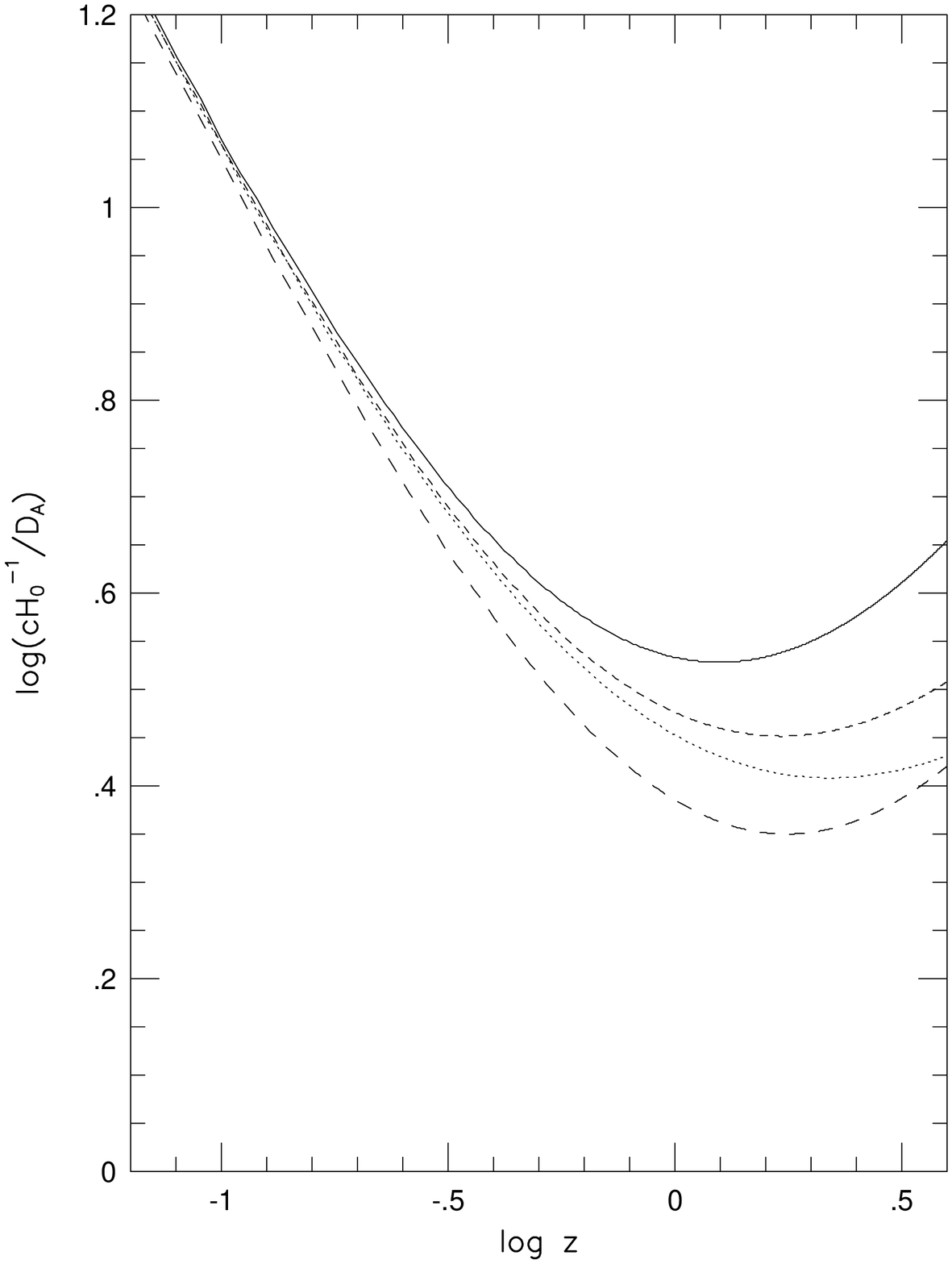}
   \caption{The $z$ dependence of the angular diameter distance $D_{\rm A}$ 
for $\alpha =
1$. Models S(1,0), O1(0.2,0), O2(0.4,0) and L(0.2,0.8) are denoted 
by solid, dotted, short dash, long dash lines, respectively. }}
\hspace{-2mm}
 \parbox{\halftext}{
\epsfxsize=7.5cm
\epsfbox{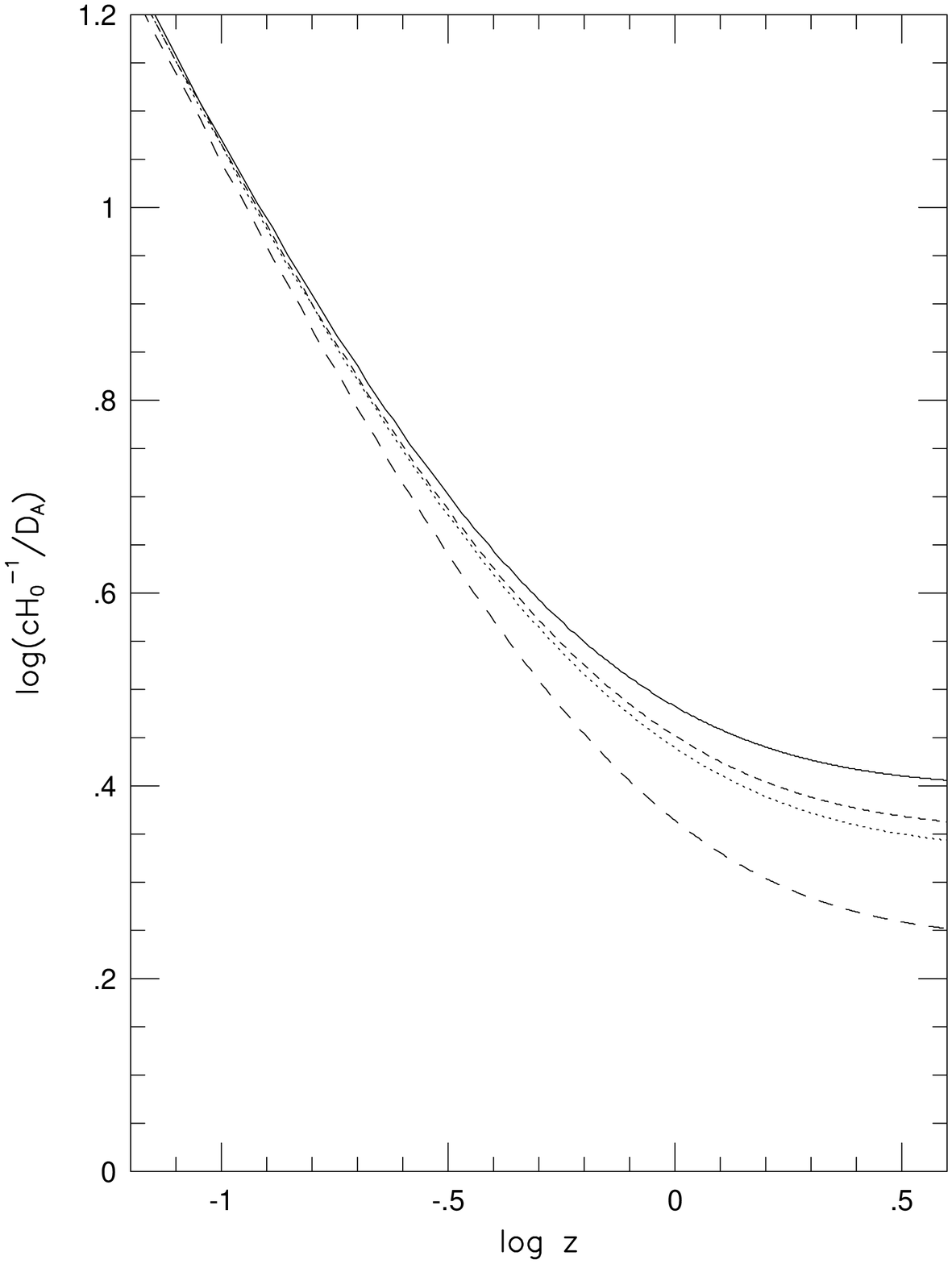}
   \caption{The $z$ dependence of the angular diameter distance $D_{\rm A}$ 
for $\alpha =
0$.  Lines have the same meaning as in Fig. 1. \vspace{0.8cm}}}
\end{figure}

\begin{figure}[htb]
 \parbox{\halftext}{
\epsfxsize=6.5cm
\epsfbox{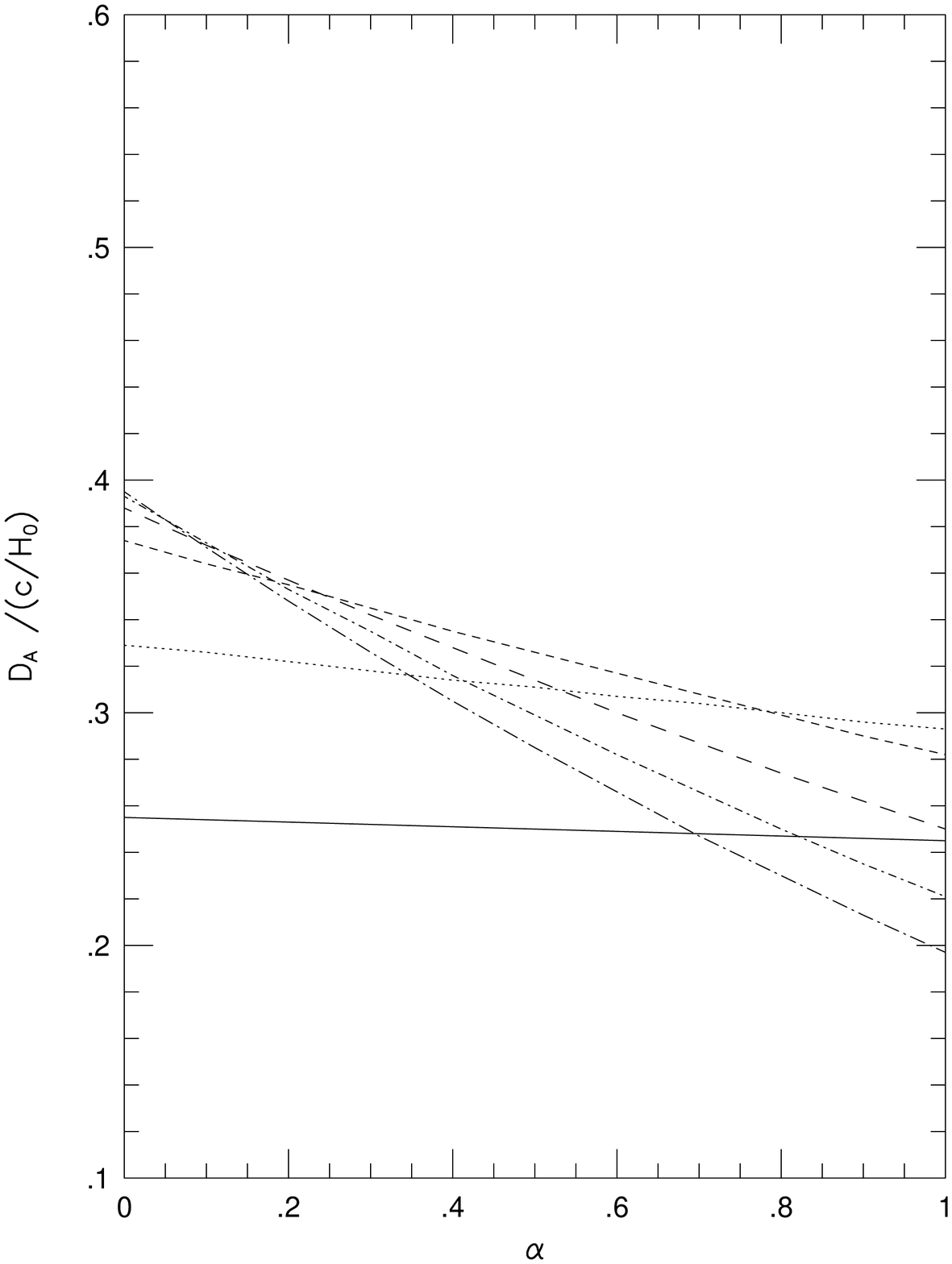}
   \caption{The $\alpha$ dependence of the angular diameter distance
$D_{\rm A}$ 
in the model S  with $(\Omega_0, \lambda_0) = (1, 0)$. The epochs 
with $z = 0.5, 1, 2, 3, 4,$ and
$5$ are denoted by solid, dotted, short dashed, long dashed lines, 
respectively. }}
\hspace{8mm}
 \parbox{\halftext}{
\epsfxsize=6.5cm
\epsfbox{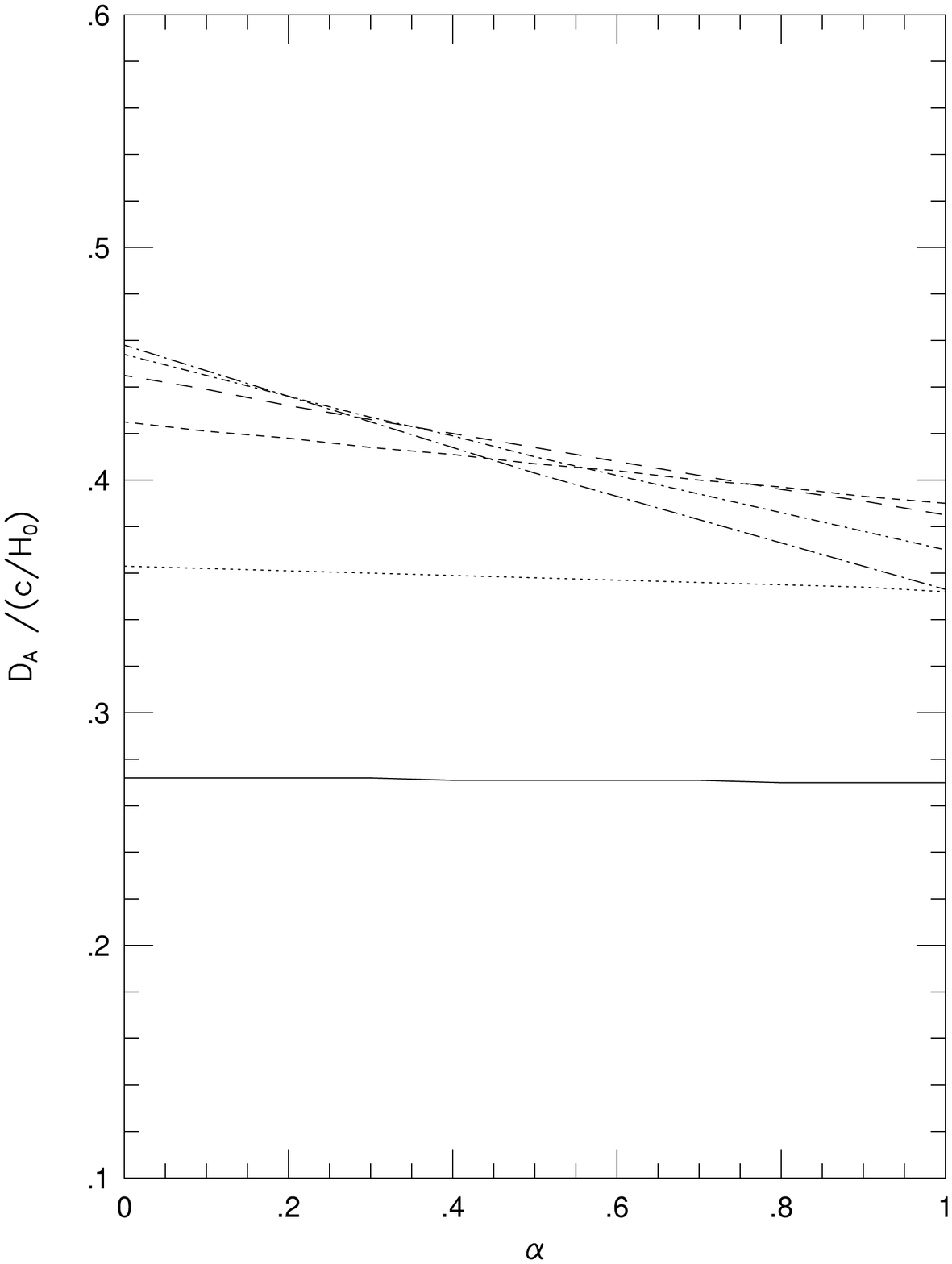}
   \caption{The $\alpha$ dependence of the angular diameter distance 
$D_{\rm A}$ 
in the model O1 with (0.2, 0).  Lines have the same meaning as in
Fig. 3.
\vspace{0.8cm}}}
\end{figure}

\begin{figure}[htb]
 \parbox{\halftext}{
\epsfxsize=6.5cm
\epsfbox{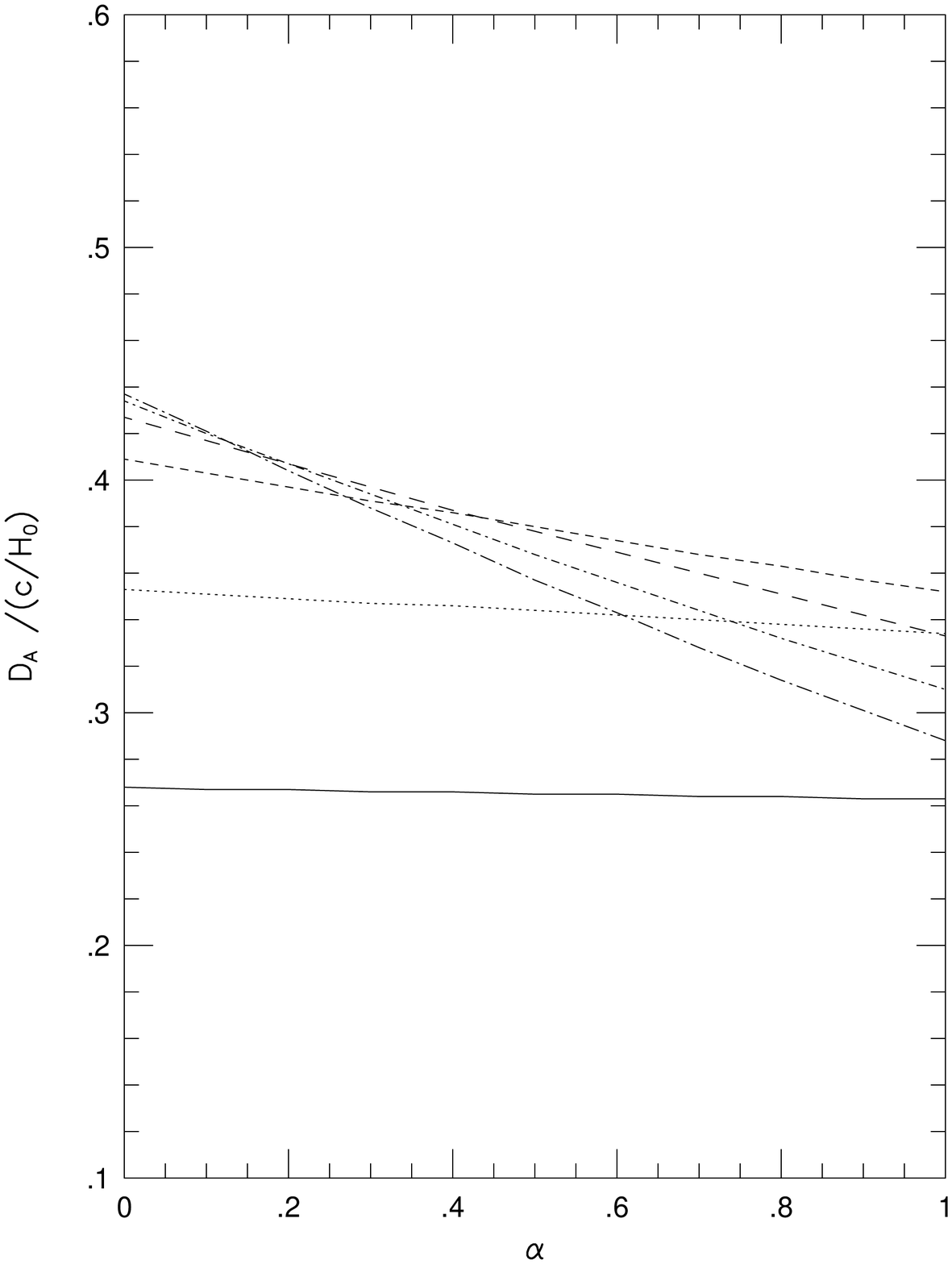}
   \caption{The $\alpha$ dependence of the angular diameter distance 
$D_{\rm A}$ 
in the model O2 with (0.4, 0).  Lines have the same meaning as in
Fig. 3.}}
\hspace{8mm} 
 \parbox{\halftext}{
\epsfxsize=6.5cm
\epsfbox{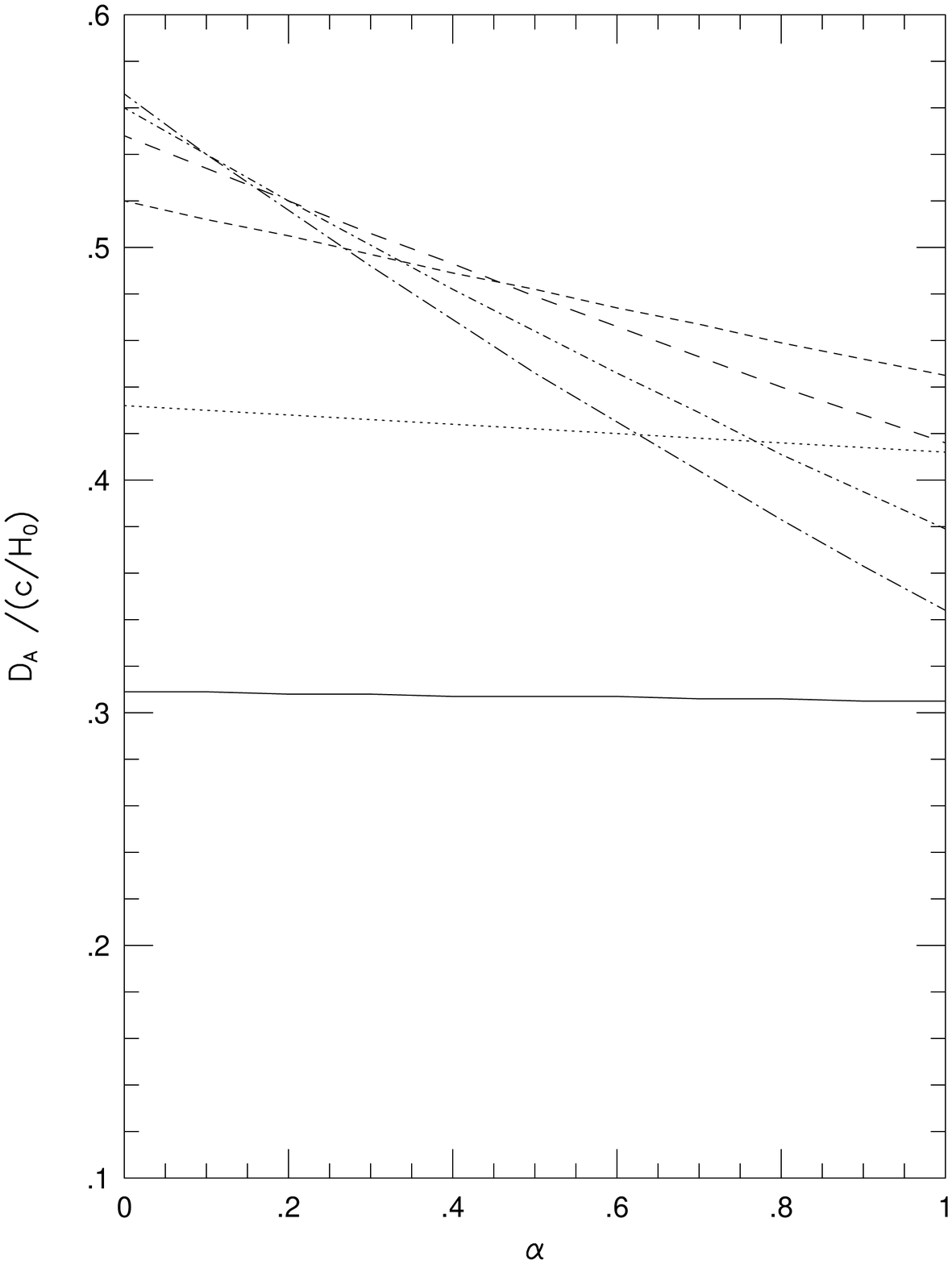}
   \caption{The $\alpha$ dependence of the angular diameter distance 
$D_{\rm A}$ 
in the model L with (0.2, 0.8).  Lines have the same meaning as in
Fig. 3.}}
\end{figure}

\bigskip

\section{Angular diameter distances in inhomogeneous models produced by 
$N$-body simulations}

For the quantitative analysis of image deformation in the lensing 
phenomena, we treated in previous papers various clumpy Friedmann models
(cf. \ Refs. 7) and 8), and also Refs. 9) and 10)).
In the present paper we use the same models for the statistical 
analysis of the angular diameter distances.
In them we considered inhomogeneities in periodic boxes
including $N = 32^3$ particles, whose distribution was
produced by performing the $N$-body simulation in the background
models S, O1, O2, and L with  $(\Omega_0, \lambda_0) = 
(1, 0), (0.2, 0), (0.4, 0),$ and $(0.2, 0.8)$, respectively. 
For the simulation we used Suto's tree code\cite{rf:su} during the
time interval $0 \leq z \leq 5$. The initial conditions were given in 
the CDM spectrum 
with the power $n = 1$ and the dispersion $\sigma_8 = 0.94$, using
Bertschinger's software {\it COSMICS},\cite{rf:bert} where the Hubble
parameter was specified as $h = 0.5, 0.7, 0.6$ and $0.7$ for model S,
O1, O2 and L, respectively.

The present sizes of the periodic boxes and the particle mass are
\begin{equation}
  \label{eq:d1}
L_0 = (32.5, 50, 39.7, 50) h^{-1} {\rm Mpc},
\end{equation}
and
\begin{equation}
  \label{eq:d2}
m \ (= \rho_{B0} {L_0}^3/N) = (2.90, 2.11, 2.11, 2.11) 
\times 10^{11} h^{-1} M_\odot
\end{equation}
for models S,  O1, O2 and L, respectively, where $\rho_{B0} = 3{H_0}^2
\Omega_0/(8\pi G)$. 

In all models, particles consist of compact lens objects which are  
galaxies and dark matter balls with the same
mass $m$ and the same constant physical radius $r_s$. The
representative value of  $r_s$ is $20 h^{-1}$ kpc, but for comparison
the cases of  $r_s 
= 10 h^{-1}$ kpc and $40 h^{-1}$ kpc are also considered. 
In a previous paper\cite{rf:t1} it was assumed that in  model S
particles consist of compact lens objects ($20 \%$) with $r_s =
20 h^{-1}$ kpc and clouds ($80 \%$) with the same mass $m$ and $r_s
\geq 20 h^{-1}$ kpc. In this paper we take the model S including only
compact lens objects with various $r_s$. Model S with larger $r_s$
corresponds effectively to the case including clouds with larger $r_s$.

The line-element of inhomogeneous models is expressed as 
\begin{equation}
  \label{eq:d3}
ds^2 = -(1+2\varphi/c^2) c^2 dt^2 
+ (1-2\varphi/c^2)a^2(t)(d\mib{x})^2/[1 + {1 \over 4} K (\mib{x})^2]^2,
\end{equation}
where  $\varphi$ is the gravitational potential connected with the
matter distribution through the Poisson equation, and $K$ is the signature 
$(\pm$ or $0)$ of the spatial curvature. The physical radius $r_s$ is
expressed using the comoving radius $x_s$ as $r_s = a(t) x_s$  
and it is taken into consideration in the form of a
softening parameter ($r_s$) in the gravitational force as 
$ 1/[a(t) (\mib{x} - \mib{x}_n)^2 + (r_s)^2]$, where $\mib{x}_n$
is the position vector of $n$-th particle.   

Let us consider a pair of rays received by the observer with the
separation angle $\theta$. By solving null-geodesic equations, which
were obtained in the previous papers,\cite{rf:t1,rf:t2}
 the interval of the two rays at any epoch can be derived.
If $(\Delta \mib{x})_\perp$ is the component of the deviation vector
perpendicular to the central direction of the rays, the angular diameter 
distance $D_{\rm A}$ is defined as
\begin{equation}
  \label{eq:d4}
D_{\rm A} = a(t) (\Delta \mib{x})_\perp [1 + {1 \over 4} K
(\mib{x})^2]^{-1}/\theta,
\end{equation}
where the factor $(1-2\varphi/c^2)$   is neglected, because
$\vert \varphi/c^2 \vert << 1$ locally. The above expression can be
rewritten by use of $y^i \ (\equiv a_0 x^i/R_0)$ as
\begin{equation}
  \label{eq:d5}
D_{\rm A} = {R_0 \over (1+z) F} (\Delta \mib{y})_\perp/\theta,
\end{equation}
where $F \equiv 1 -{1 \over 4}(R_0H_0/c)^2 (1 -\Omega_0 -\lambda_0) 
(\mib{y})^2, \ a_0 = a(t_0), $ and \ $R_0 \equiv L_0/N^{1/3}$.

Behavior of the angular diameter distance depends on the separation angle
$\theta$. In the cosmological observation for angular sizes of compact 
radio sources by Kellerman,\cite{rf:kel,rf:jd} it was found to be on the 
order of milli-arcseconds, or $\theta \sim
0.005$ arcsec. Accordingly we adopt $\theta = 0.005$ arcsec as a 
representative value and for comparison $\theta = 0.1$ and $20$ arcsec
as other representative values.

In the present lensing simulation we considered five parameter sets 
$(r_s, \theta)$, \ A. ($20 h^{-1}$kpc, 0.005 arcsec), \ B. ($40 h^{-1}
$kpc, 0.005 arcsec), \ C. ($10 h^{-1}$kpc, 0.005 arcsec), \
D. ($20 h^{-1}$kpc, 0.1 arcsec) and E. ($20 h^{-1}$kpc, 20 arcsec), 
and performed the ray-shooting of 500 
ray pairs for each parameter set in the four models. At the six epochs 
$z = 0.5, 1, 2, 3, 4$ and $5$, we compared the calculated distances
with the Dyer-Roeder distance and determined the corresponding value 
of $\alpha$. Since the angular diameter distance depends on $\alpha$
linearly for $0 \leq z \leq 5$, we define  $\alpha$ as follows for the
calculated distance $D_{\rm A}$ :
\begin{equation}
  \label{eq:d6}
\alpha = (D_{\rm A} - D_{\rm DR})/(D_{\rm F} - D_{\rm DR}),
\end{equation}
where $D_{\rm DR}$ is the limiting Dyer-Roeder distance with $\alpha = 0$
and $D_{\rm F}$ is the calculated Friedmann distance in the homogeneous
case. This $D_{\rm F}$ is equal to the Dyer-Roeder distance with 
$\alpha = 1$, except for small errors in numerical integrations.

\begin{figure}[htb]
 \parbox{\halftext}{
\epsfxsize=7.5cm
\epsfbox{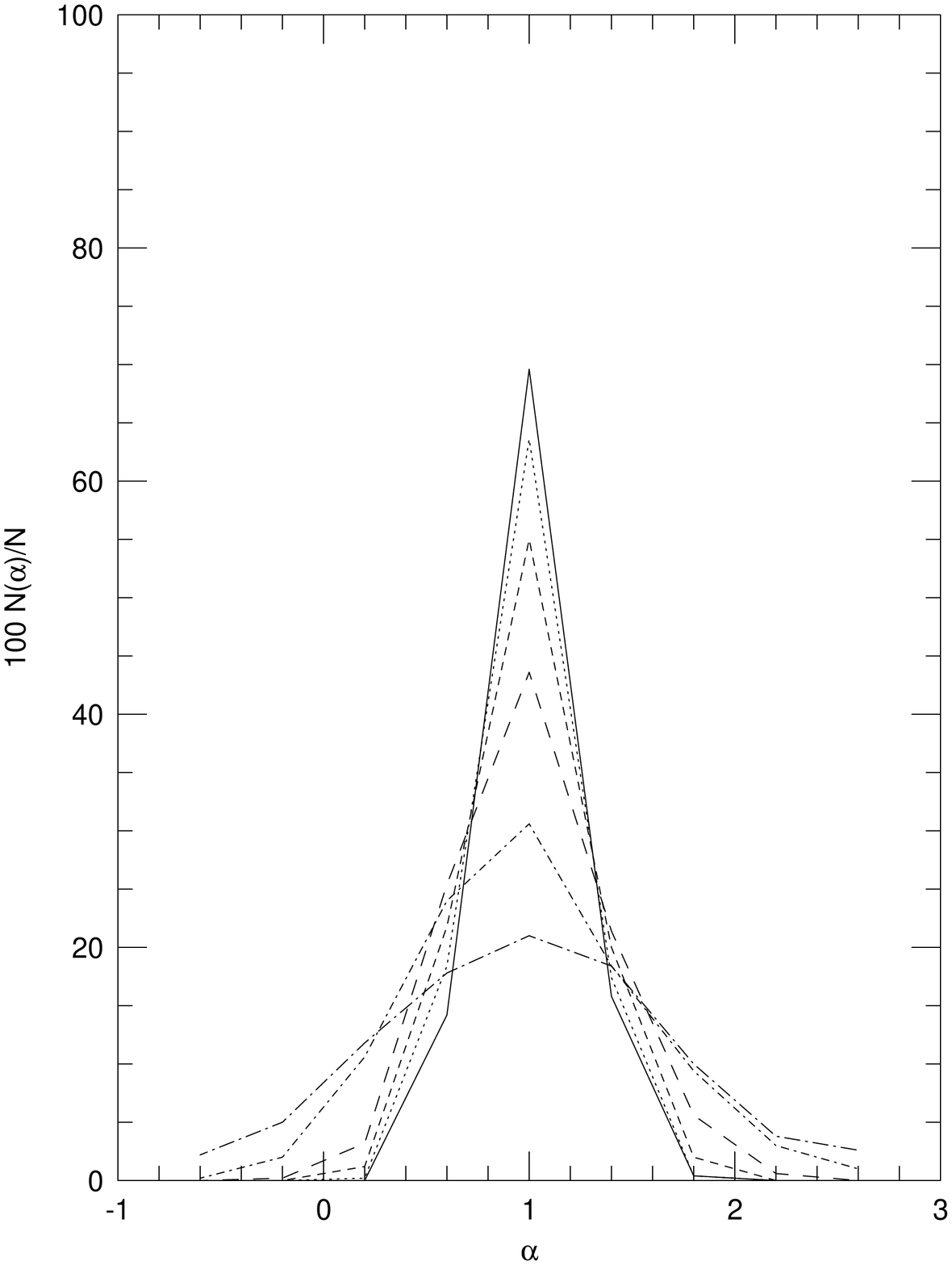}
   \caption{The percentage ($100 N(\alpha)/ N)$ of the distribution 
of $\alpha$ in bins with the interval $\Delta \alpha = 0.4$, for the
parameter set A(20, 0.005) in model S with $(\Omega_0, \lambda_0) =
(1.0, 0)$. Results for $z = 0.5, 1, 2, 3, 4, 5$ are denoted
by dot-long dashed, dot-short dashed, long dashed, short dashed, 
dotted, solid lines, respectively}}
\hspace{-2mm}
 \parbox{\halftext}{
\epsfxsize=7.5cm
\epsfbox{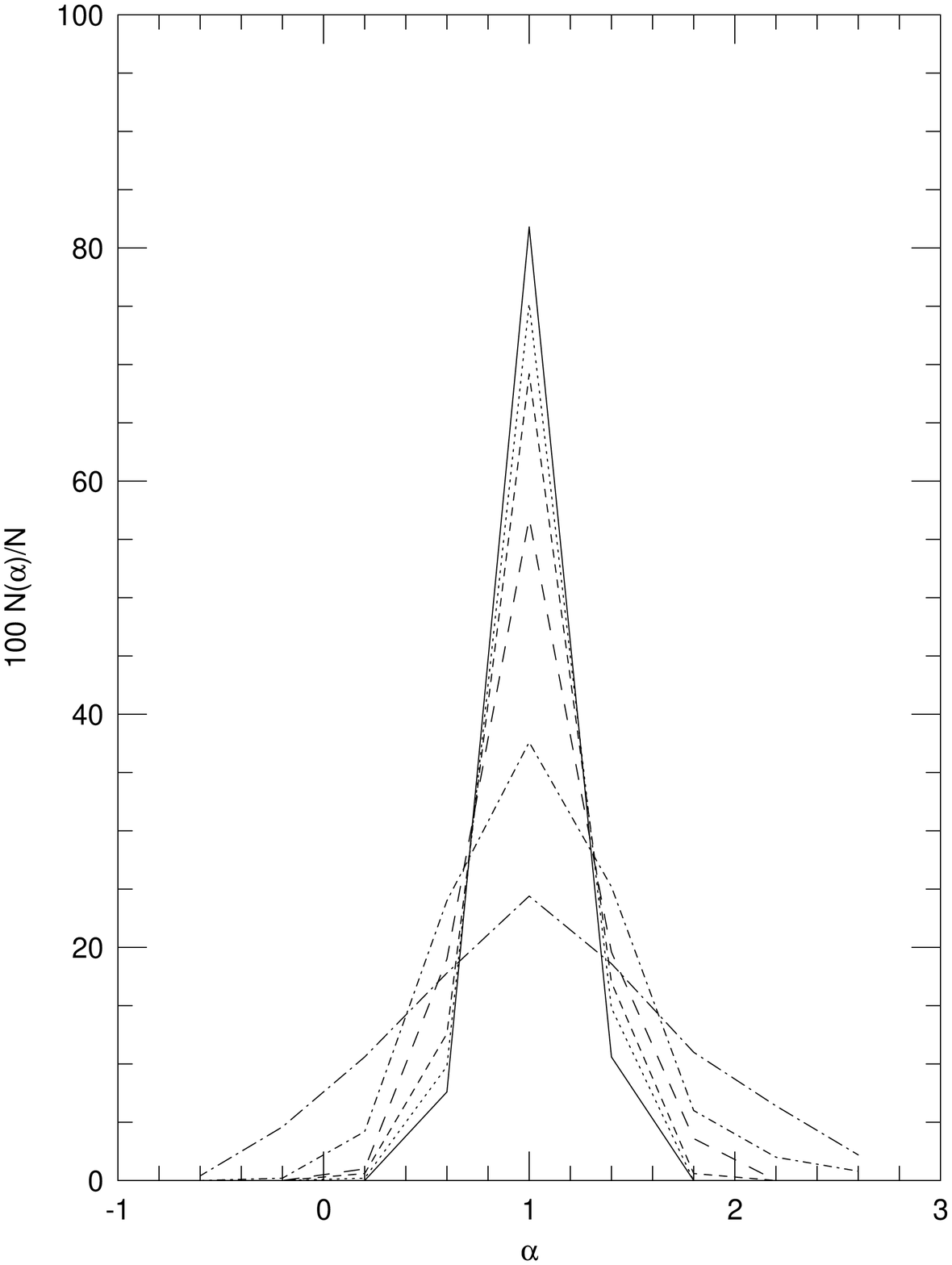}
   \caption{The percentage ($100 N(\alpha)/ N)$ of the distribution 
of $\alpha$ in bins with the interval $\Delta \alpha = 0.4$, for the
parameter set B(40, 0.005) in model S with $(1.0, 0)$.
  Lines have the same meaning as in Fig. 7. \vspace{0.8cm}}}
\end{figure}

\begin{figure}[htb]
 \parbox{\halftext}{
\epsfxsize=7.5cm
\epsfbox{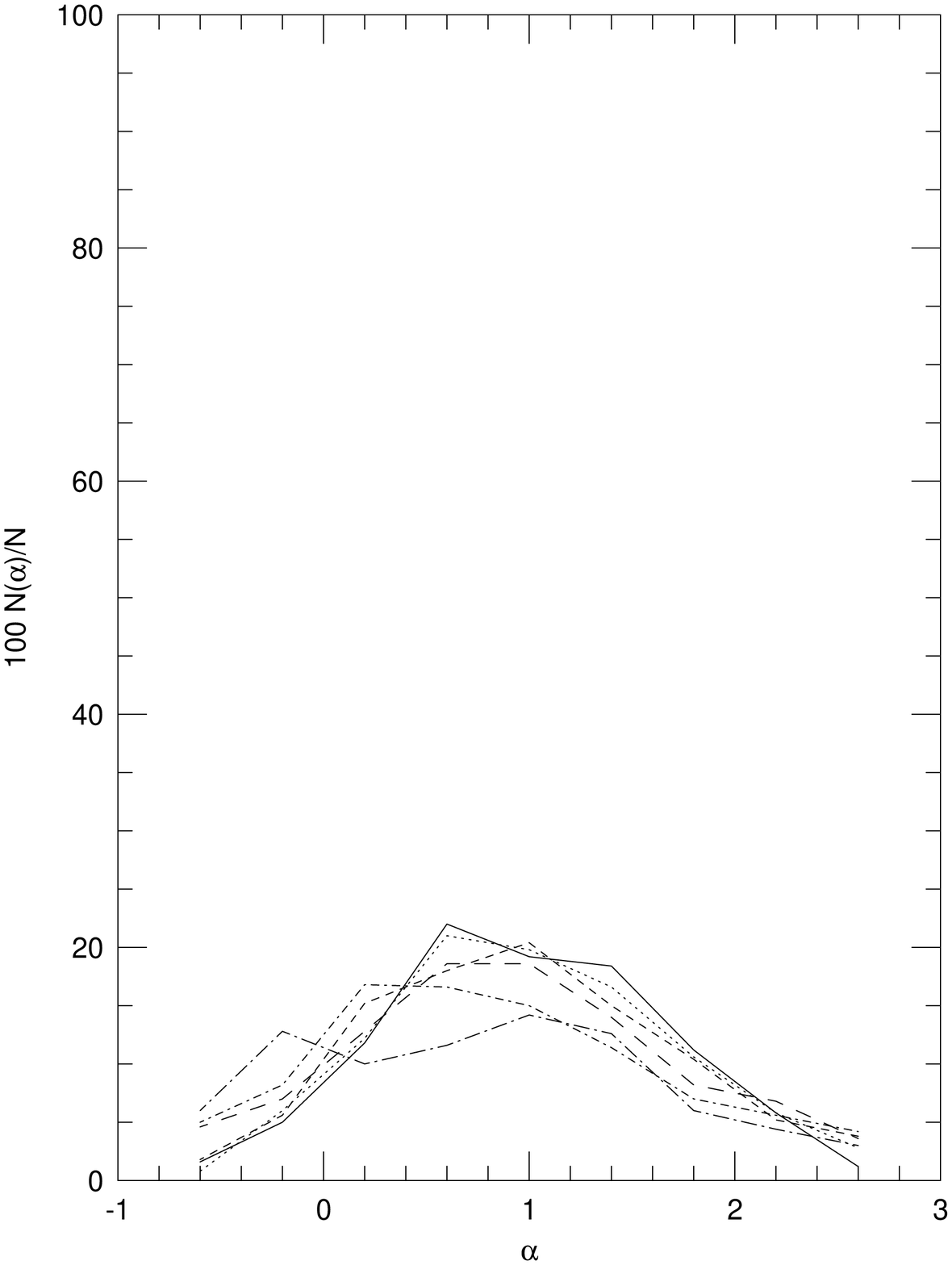}
   \caption{The percentage ($100 N(\alpha)/ N)$ of the distribution 
of $\alpha$ in bins with the interval $\Delta \alpha = 0.4$, for the
parameter set A(20, 0.005) in model O1 with $(0.2, 0)$. 
Lines have the same meaning as in Fig. 7. }}
\hspace{-2mm}
 \parbox{\halftext}{
\epsfxsize=7.5cm
\epsfbox{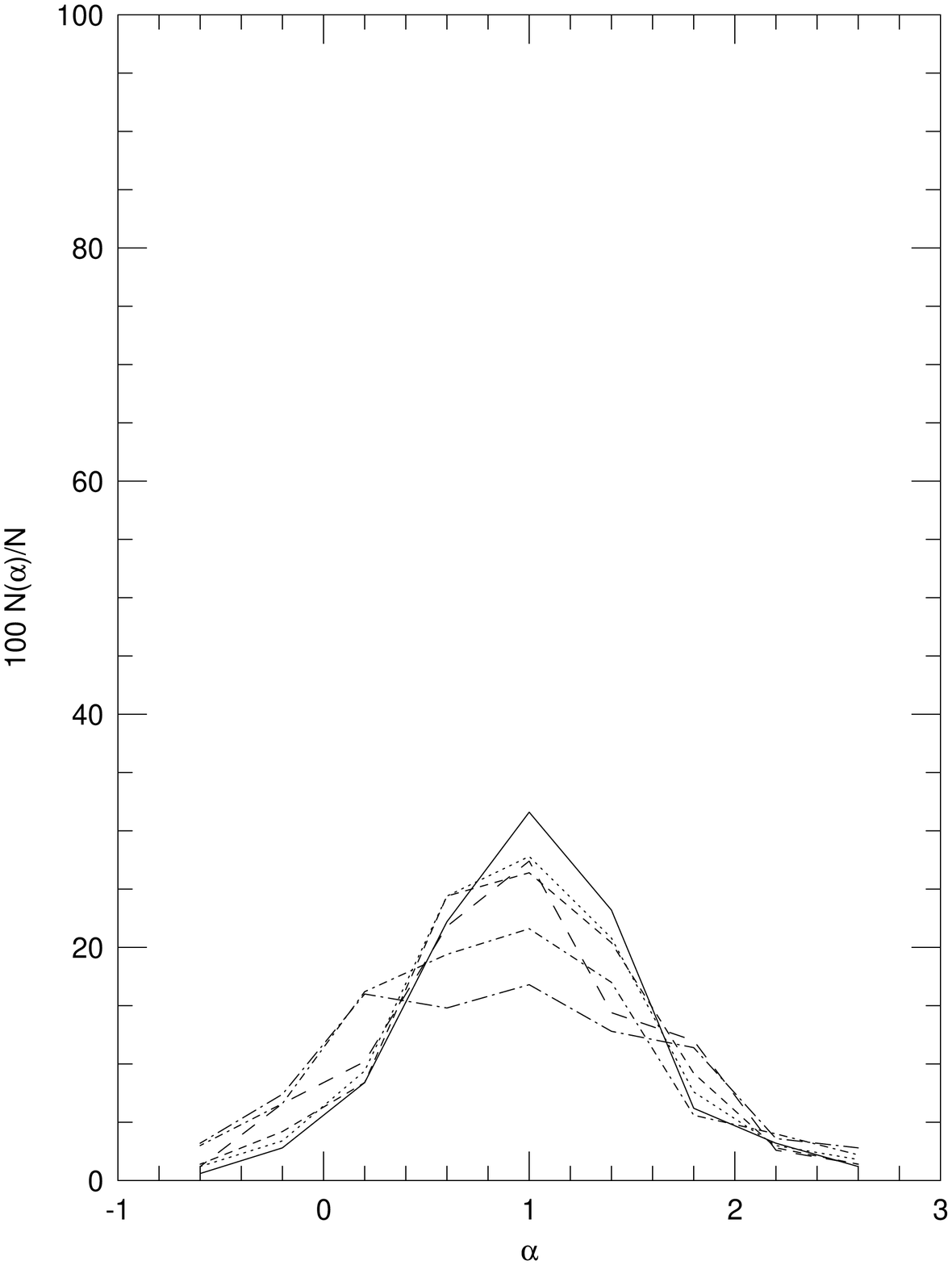}
   \caption{The percentage ($100 N(\alpha)/ N)$ of the distribution 
of $\alpha$ in bins with the interval $\Delta \alpha = 0.4$, for the
parameter set B(40, 0.005) in model O1 with (0.2, 0).
  Lines have the same meaning as in Fig. 7. }}
\end{figure}
   
\begin{figure}[htb]
 \parbox{\halftext}{
\epsfxsize=7.5cm
\epsfbox{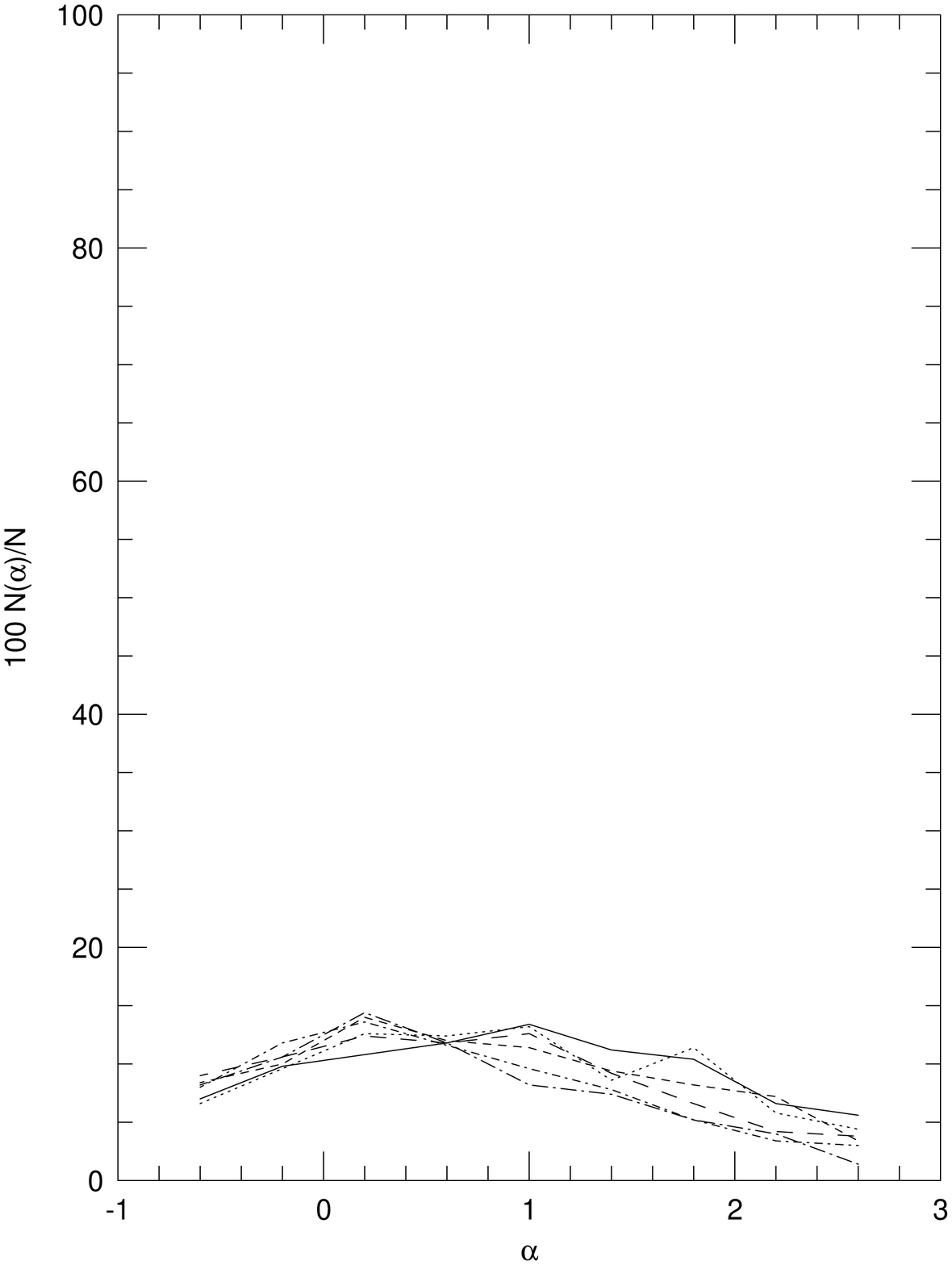}
   \caption{The percentage ($100 N(\alpha)/ N)$ of the distribution 
of $\alpha$ in bins with the interval $\Delta \alpha = 0.4$, for the
parameter set C(10, 0.005) in model O1 with $(0.2, 0)$. 
Lines have the same meaning as in Fig. 7. }}
\hspace{-2mm}
 \parbox{\halftext}{
\epsfxsize=7.5cm
\epsfbox{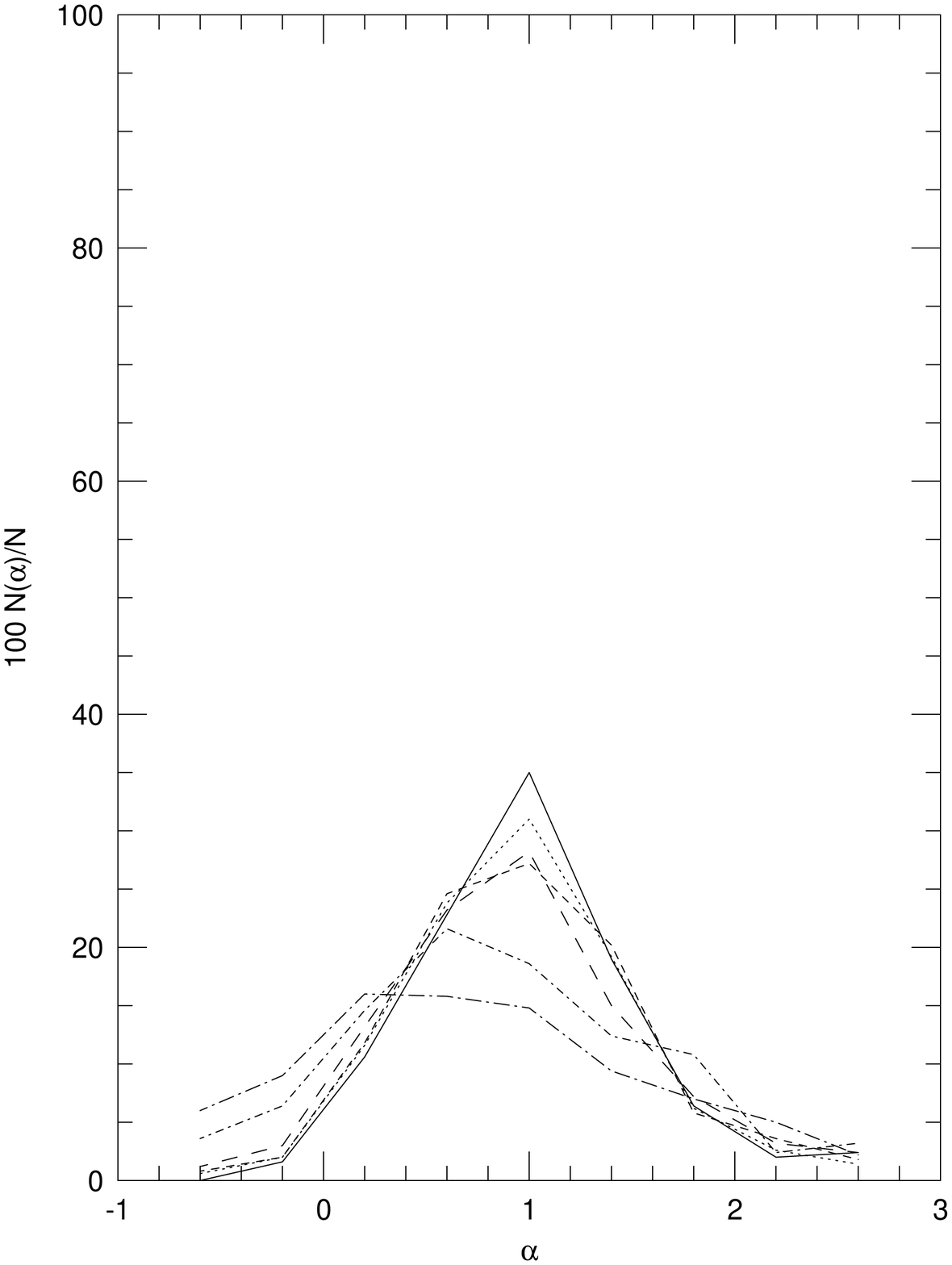}
   \caption{The percentage ($100 N(\alpha)/ N)$ of the distribution 
of $\alpha$ in bins with the interval $\Delta \alpha = 0.4$, for the
parameter set E(20, 20) in model O1 with (0.2, 0).
  Lines have the same meaning as in Fig. 7. }}
\end{figure}

\begin{figure}[htb]
 \parbox{\halftext}{
\epsfxsize=7.5cm
\epsfbox{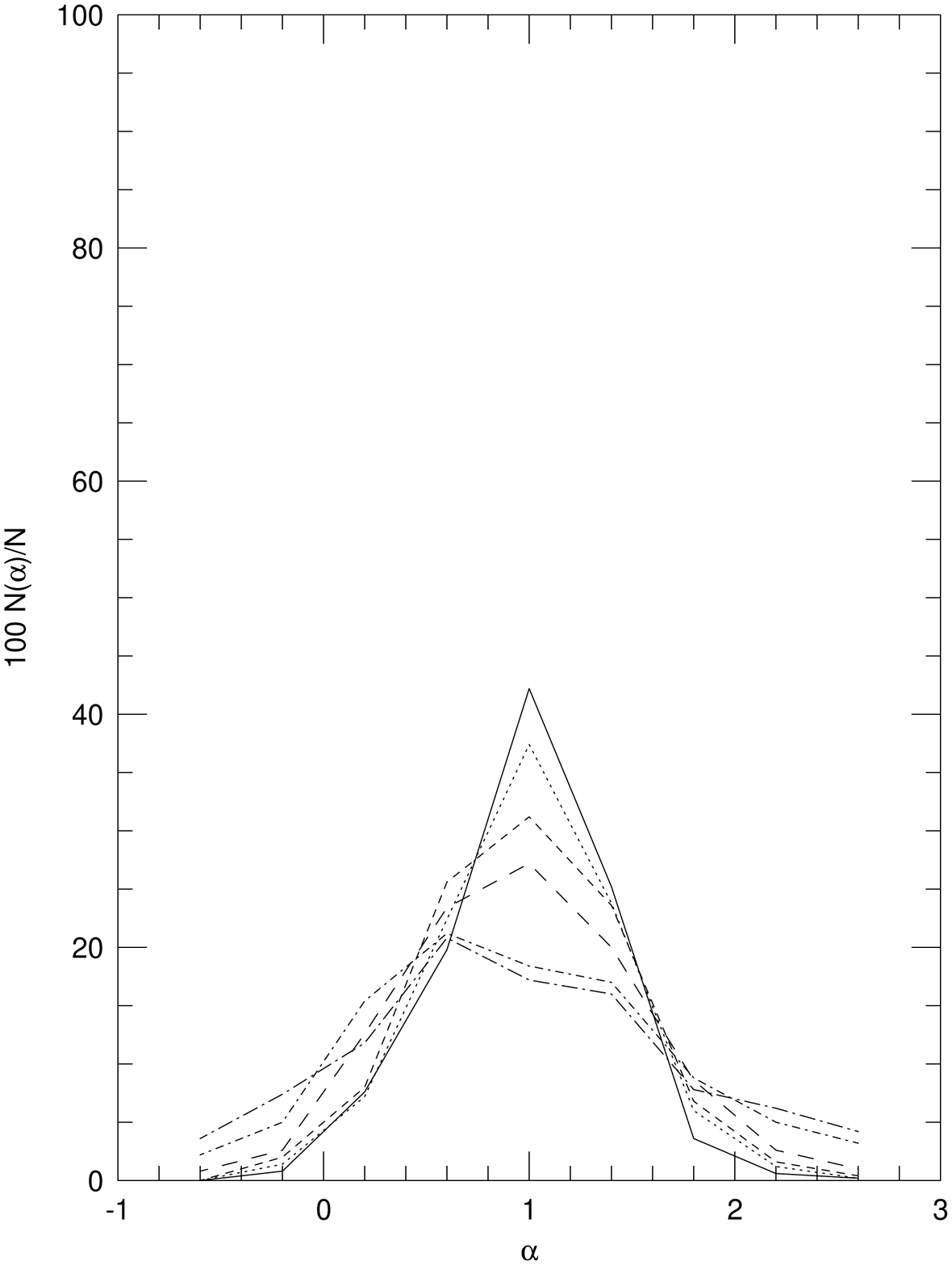}
   \caption{The percentage ($100 N(\alpha)/ N)$ of the distribution 
of $\alpha$ in bins with the interval $\Delta \alpha = 0.4$, for the
parameter set A(20, 0.005) in model O2 with $(0.4, 0)$. 
Lines have the same meaning as in Fig. 7. }}
\hspace{-2mm}
 \parbox{\halftext}{
\epsfxsize=7.5cm
\epsfbox{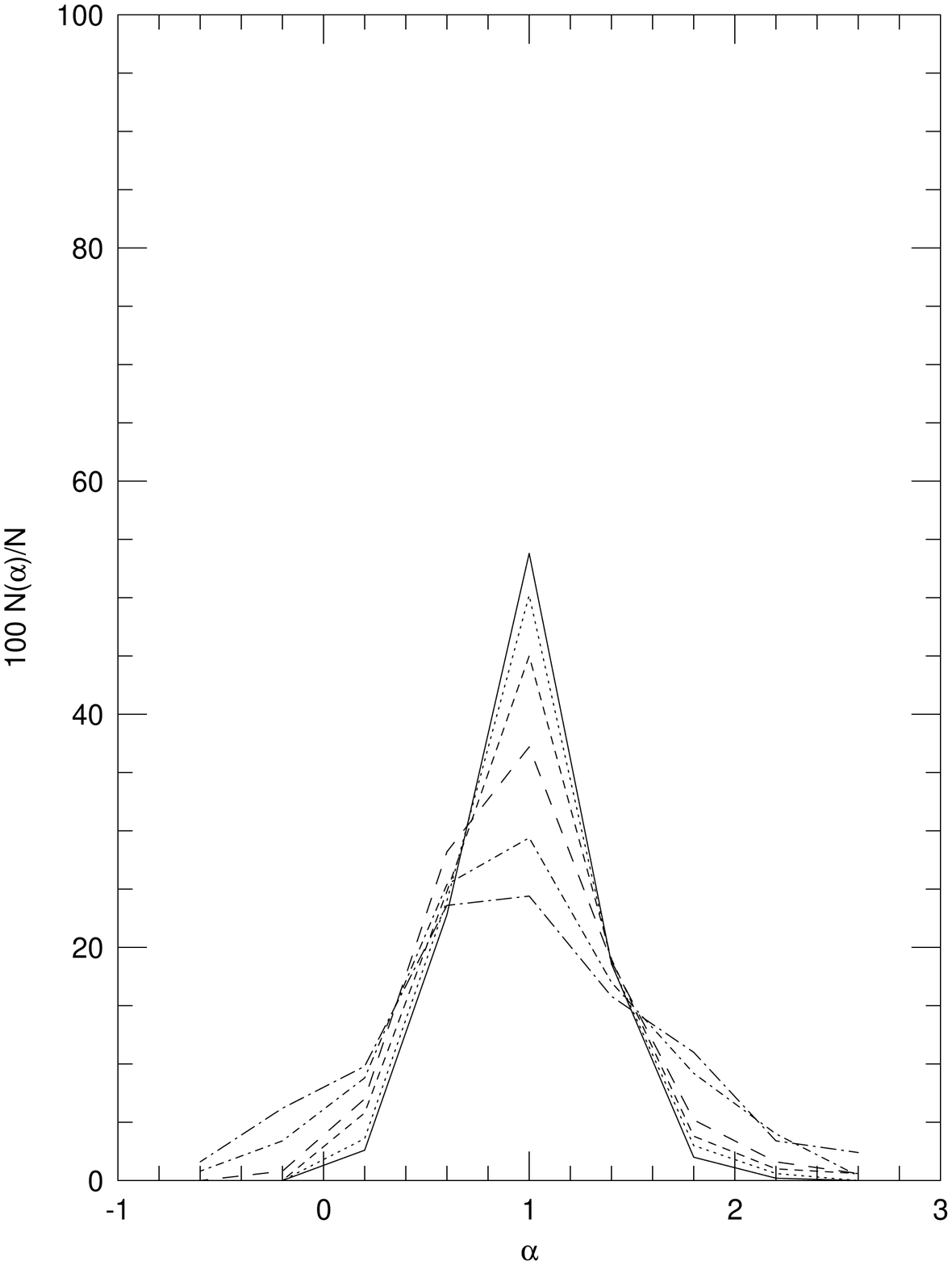}
   \caption{The percentage ($100 N(\alpha)/ N)$ of the distribution 
of $\alpha$ in bins with the interval $\Delta \alpha = 0.4$, for the
parameter set B(40, 0.005) in model O2 with (0.4, 0).
  Lines have the same meaning as in Fig. 7. }}
\end{figure}

\begin{figure}[htb]
 \parbox{\halftext}{
\epsfxsize=7.5cm
\epsfbox{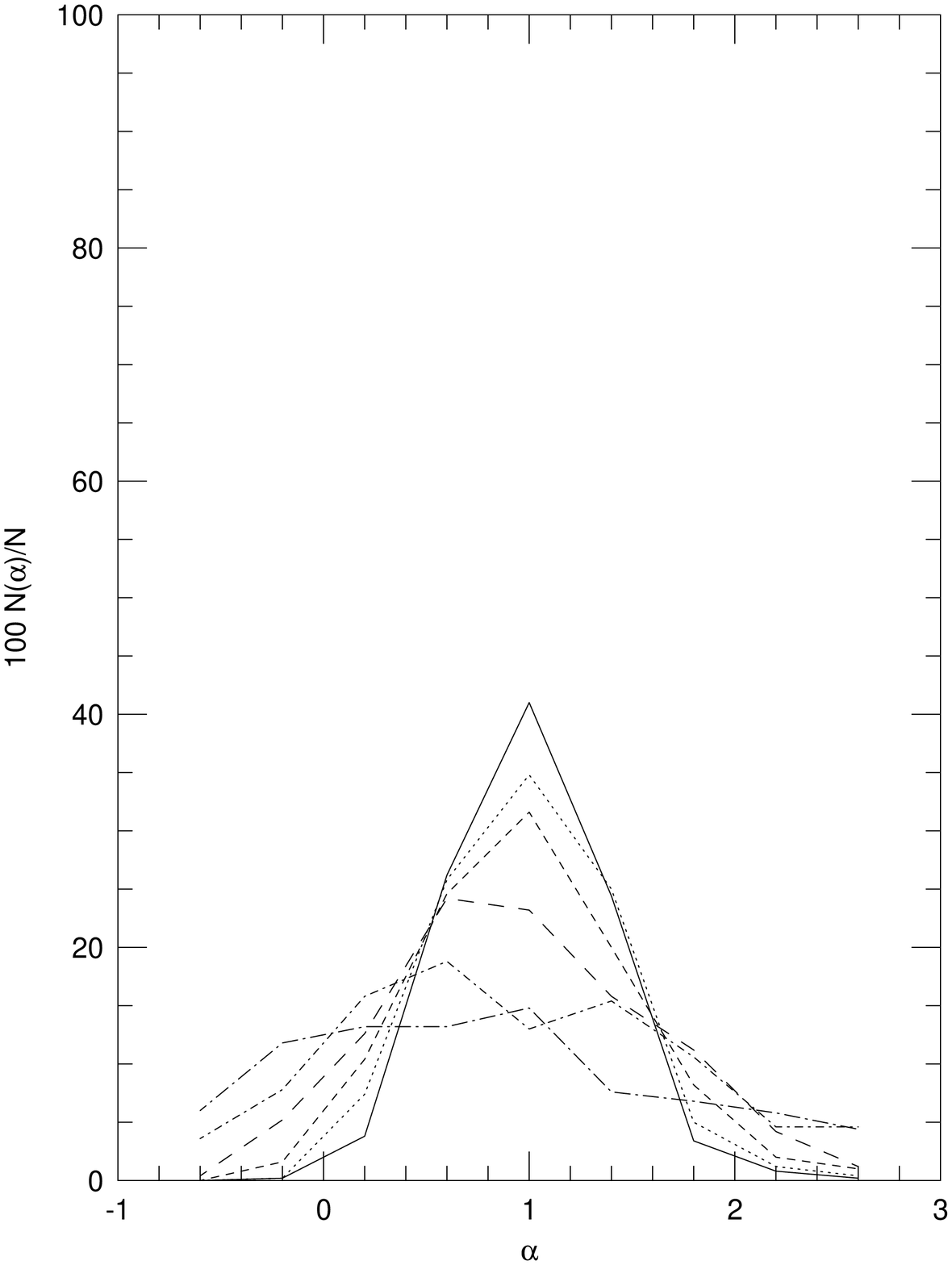}
   \caption{The percentage ($100 N(\alpha)/ N)$ of the distribution 
of $\alpha$ in bins with the interval $\Delta \alpha = 0.4$, for the
parameter set A(20, 0.005) in model L with $(0.2, 0.8)$. 
Lines have the same meaning as in Fig. 7. }}
\hspace{-2mm}
 \parbox{\halftext}{
\epsfxsize=7.5cm
\epsfbox{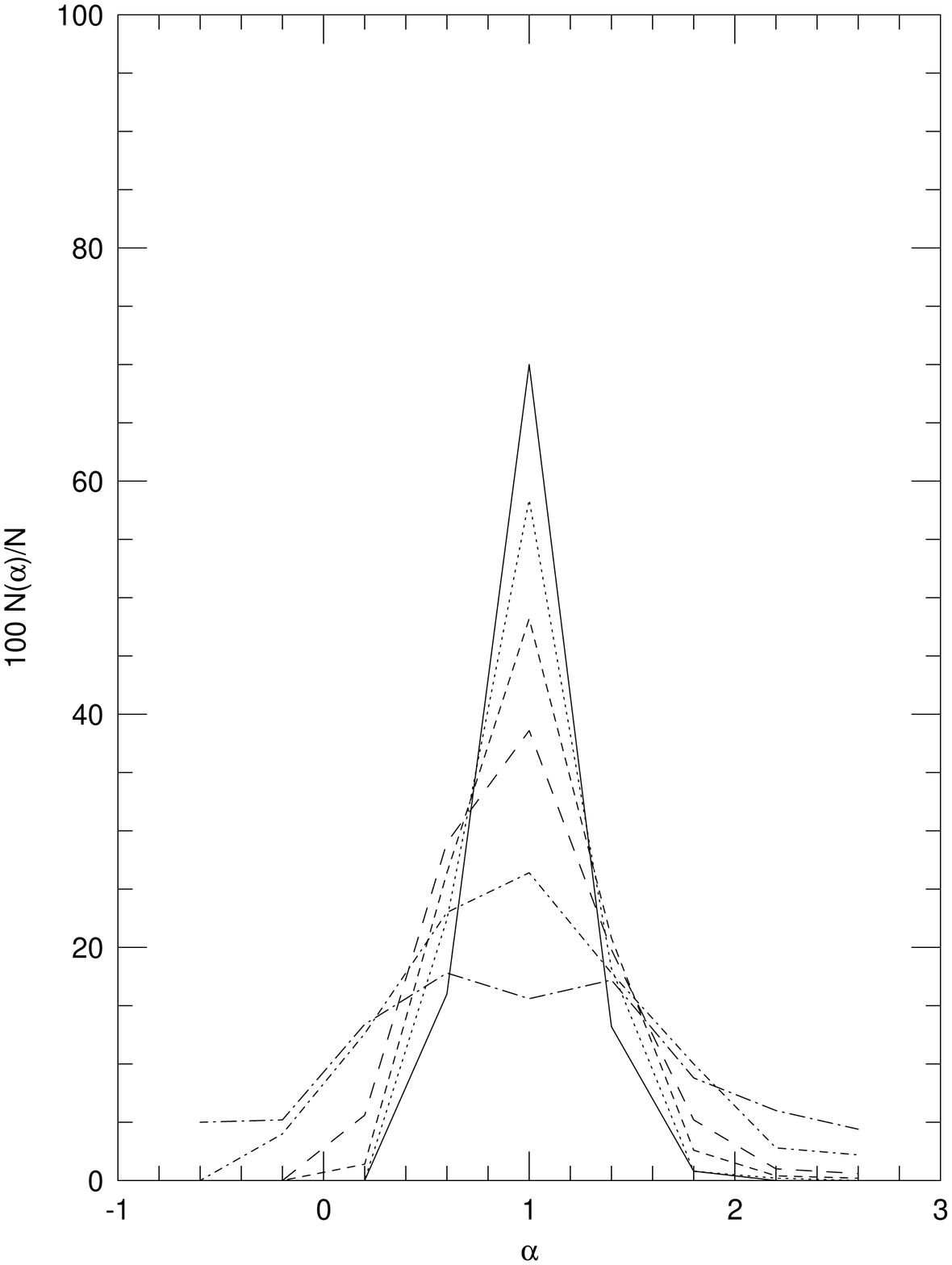}
   \caption{The percentage ($100 N(\alpha)/ N)$ of the distribution 
of $\alpha$ in bins with the interval $\Delta \alpha = 0.4$, for the
parameter set B(40, 0.005) in model L with (0.2, 0.8).
  Lines have the same meaning as in Fig. 7. }}
\end{figure}
   
\begin{figure}[htb]
 \parbox{\halftext}{
\epsfxsize=7.5cm
\epsfbox{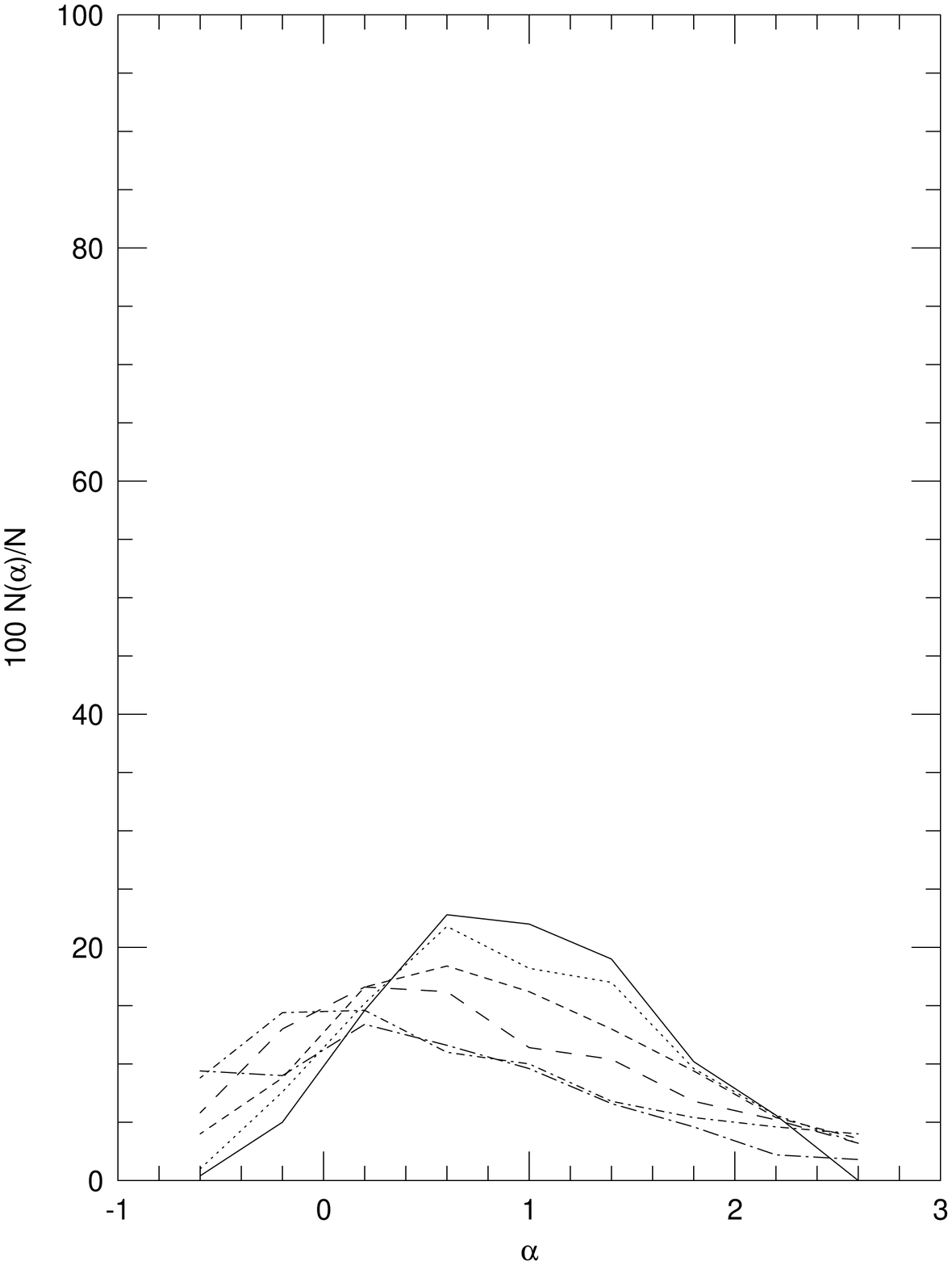}
   \caption{The percentage ($100 N(\alpha)/ N)$ of the distribution 
of $\alpha$ in bins with the interval $\Delta \alpha = 0.4$, for the
parameter set C(10, 0.005) in model L with $(0.2, 0.8)$. 
Lines have the same meaning as in Fig. 7. }}
\hspace{-2mm}
 \parbox{\halftext}{
\epsfxsize=7.5cm
\epsfbox{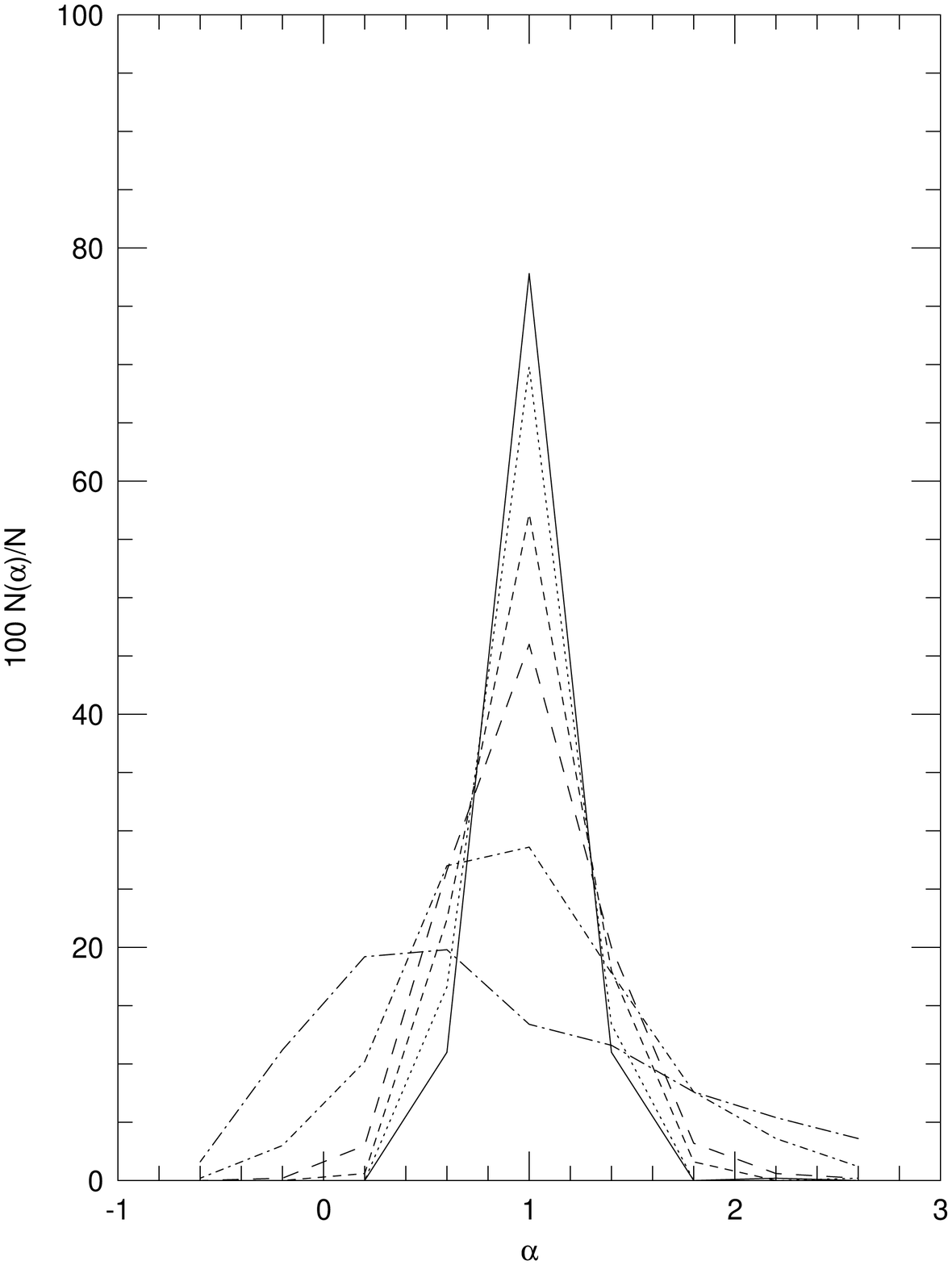}
   \caption{The percentage ($100 N(\alpha)/ N)$ of the distribution 
of $\alpha$ in bins with the interval $\Delta \alpha = 0.4$, for the
parameter set E(20, 20) in model L with (0.2, 0.8).
  Lines have the same meaning as in Fig. 7. }}
\end{figure}

As a result of statistical analysis for this ray-shooting, we
derived the average clumpiness parameter $\bar{\alpha}$, the
dispersion  $\sigma_\alpha$, and the distribution ($N(\alpha)$) of
$\alpha$. Here in order to study the frequency of ray pairs with
$\alpha$, we consider many bins with the interval $\Delta \alpha =
0.4$ and the centers $\alpha_i = 1.0 \pm 0.4 i \ (i = 0, 1, 2, ...)$.
The number of ray pairs with $\alpha_i - \Delta \alpha/2 \leq \alpha 
\leq \alpha_i + \Delta \alpha/2$ is $N (\alpha_i)$ and the total
number of ray pairs is $N (= \sum_i N(\alpha_i))$.  
In Tables I $\sim$ IV we show $(\bar{\alpha}, \sigma_\alpha)$ for five
parameters sets in model S, O1, O2 and L, respectively. In Figs. 7 $\sim$
18, we show the percentages of the distribution of $\alpha$, that is, 
$100 N(\alpha)/ N$ for several parameter sets.
The following types of statistical behavior are found from these 
tables and figures :

\noindent (1) $\bar{\alpha}$ is nearly equal to $1$ in all
models. However the individual values differ considerably from  $1$ 
for $\theta < 1$ arcsec, because the dispersion $\sigma_\alpha$ is 
large enough (for example $\sigma_\alpha \sim 0.5$ and $1$ for $z = 
1 - 2$ in model S and low-density models, respectively). Of the four
models the dispersions are smallest and largest in models S and O1,
respectively. Those in O2 and L are comparable. This situation is 
reflected also in the behaviors of the distribution of $\alpha$, 
that is, around the point $\alpha = 1$ it is most symmetric and 
asymmetric in models S and O1, respectively.  

\noindent (2) The dependence of $\sigma_\alpha$ on the radius of 
particles is large, $\sigma_\alpha$ decreases with an increase of
$r_s$, and when $r_s << 20 h^{-1}$kpc, the dispersion $\sigma_\alpha$ is 
too large to determine $\bar{\alpha}$. The symmetry of the 
distribution of $\alpha$ increases in
the order of C, A, and B (for the same $\theta$) in all models.
This is because with an increase of $r_s$ the volume of empty regions 
included in the separation angles of ray pairs decreases and the 
inhomogeneous matter distribution in them is smoothed.

\noindent (3) With the increase of the redshift $z$, the dispersion 
$\sigma_\alpha$ decreases and the symmetry of the distribution 
increases in all models. This is because with an increase of $z$
the number of particles included within the separation angles increases
and the gravitational effect on ray pairs is homogenized.  

\noindent (4) For $\theta \leq 1$ arcsec, the dependence of $\sigma_\alpha$ 
on the separation angle $\theta$ is small, as can be found by comparing
columns A and D of the tables. In low-density models the
dependence is rather large for $\theta >> 1$ arcsec, as in column
E.  This is because in low-density models the quantity of matter
included within the separation angle of ray pairs is much different in 
the cases of $\theta \leq 1$ and $\theta >> 1$ arcsec, and the corresponding
distance for $\theta >>1$ arcsec is clearly close to
the Friedmann distance, in contrast to the case of  $\theta \leq 1$
arcsec. 

\noindent (5) In the case of $\sigma_\alpha \geq 0.8$, the
distribution of $\alpha$
is irregular and asymmetric around $\alpha = 1$ and generally $N(\alpha)$ 
for $\alpha < 1$ is larger than that for $\alpha > 1$ (cf. Figs. 9 $\sim$
11, 13 $\sim$ 17). This behavior suggests that the distribution function of 
$\alpha$ is not Gaussian in these cases. For ray pairs with 
$\sigma_\alpha \geq 0.8$, the $z$ dependence of $\bar{\alpha}$
also is irregular and may decrease slightly from $1$ with an increase 
of $z$. For ray pairs with $\sigma_\alpha < 0.5$, however,
$\bar{\alpha}$ clearly approaches $1$ with an increase of $z$.

\noindent (6) A positive cosmological constant has the role of
decreasing $\sigma_\alpha$ and the asymmetry of the distribution of
$\alpha$  (cf. Figs. 9 $\sim$ 12 and Figs. 15 $\sim$ 18).

 Figures 3 $\sim$ 6 in the previous section show for $z \sim 0.5$ that the 
change in the Dyer-Roeder distance ($D_{\rm A}$) for the interval
$\alpha = 0 \sim 1$ is very small. Accordingly, we can approximately 
use  $D_{\rm A}$ with $\alpha = 1$ for $z \sim 0.5$, even if 
$\sigma_\alpha$ is larger than 1. For $z > 1$ and $\theta < 1$ arcsec, 
however, the change in
$D_{\rm A}$ for this interval of $\alpha$ is not so small that the large
dispersion ($\sigma_\alpha$) brings a considerable fluctuation in
the value of $D_{\rm A}$. This means that for $z >
1$ and $\theta < 1$ arcsec, the realistic disntance may deviate
considerably from the Friedmann distance in low-density models.

\begin{table}
\caption{The average clumpiness parameter $\bar{\alpha}$ and its dispersion 
$\sigma_\alpha$ in model S with $(\Omega_0, \lambda_0) = (1.0, 0)$. \
Columns A, B, C, D and E correspond to the parameter sets (20, 0.005), (40,
0.005), (10, 0.005), (20, 0.1) and (20, 20), respectively.}  
\label{table:1}
\begin{center}
\begin{tabular}{crcrcrcrcrc} \hline \hline
 & A & & B & & C & & D & & E
\\ \hline
$z$&$\bar{\alpha} \quad$ & $\sigma_\alpha$ & $\bar{\alpha}
 \quad $ & $\sigma_\alpha$
&$\bar{\alpha} \quad $ & $\sigma_\alpha$ &$\bar{\alpha} \quad $ & $\sigma_\alpha$
&$\bar{\alpha} \quad $ & $\sigma_\alpha$ 
\\ \hline
0.5& $1.07$ &  $0.93$ &  1.12 &  0.78 & $1.01$ &  1.31& $1.08$ &  0.84 
 &1.06 & 0.76\\
1&   $1.01$ &  $0.57$ &  1.05 &  0.43 & $0.98$ &  0.81& $1.02$ &  0.50 
 &1.01 & 0.40\\
2&   $1.01$ &  $0.36$ &  1.02 &  0.28 & $0.99$ &  0.54& $1.01$ &  0.33 
 &1.00 & 0.26\\
3&   $1.00$ &  $0.27$ &  1.02 &  0.21 & $0.99$ &  0.42& $1.01$ &  0.25 
 &1.00 & 0.19\\
4&   $1.00$ &  $0.23$ &  1.01 &  0.18 & $0.99$ &  0.35& $1.01$ &  0.21
 &1.00 & 0.16\\
5&   $1.00$ &  $0.20$ &  1.01 &  0.16 & $0.99$ &  0.30& $1.00$ &  0.18
 &1.00 & 0.14\\ 
\hline
\end{tabular}
\end{center}
\end{table}
\bigskip

\begin{table}
\caption{The average clumpiness parameter $\bar{\alpha}$ and its dispersion 
$\sigma_\alpha$ in model O1 with (0.2, 0).} 
\label{table:2}
\begin{center}
\begin{tabular}{crcrcrcrcrc} \hline \hline
 & A & & B & & C & & D & & E
\\ \hline
$z$&$\bar{\alpha} \quad$ & $\sigma_\alpha$ & $\bar{\alpha}
 \quad $ & $\sigma_\alpha$
&$\bar{\alpha} \quad $ & $\sigma_\alpha$ &$\bar{\alpha} \quad $ 
& $\sigma_\alpha$ &$\bar{\alpha} \quad $ & $\sigma_\alpha$ 
\\ \hline
0.5& $1.01$ &  $2.42$ &  1.23 &  1.28 & $1.07$ & 4.08& $0.99$ & 2.39 
 &0.91 & 1.48\\
1&   $0.94$ &  $1.43$ &  1.05 &  0.86 & $0.91$ & 2.38& $0.93$ & 1.42
 &0.96 & 1.08\\
2&   $1.00$ &  $1.03$ &  1.02 &  0.73 & $0.95$ & 1.63& $0.99$ & 0.99
 &1.00 & 0.78\\
3&   $0.98$ &  $0.89$ &  1.01 &  0.63 & $0.95$ & 1.37& $0.99$ & 0.87
 &1.00 & 0.66\\
4&   $0.98$ &  $0.85$ &  0.99 &  0.59 & $0.93$ & 1.31& $0.98$ & 0.83
 &1.00 & 0.60\\
5&   $0.96$ &  $0.80$ &  0.97 &  0.55 & $0.89$ & 1.26& $0.97$ & 0.78
 &1.00 & 0.54\\ 
\hline
\end{tabular}
\end{center}
\end{table}
\bigskip

\begin{table}
\caption{The average clumpiness parameter $\bar{\alpha}$ and its dispersion 
$\sigma_\alpha$ in model O2 with (0.4, 0). } 
\label{table:3}
\begin{center}
\begin{tabular}{crcrcrcrcrc} \hline \hline
 & A & & B & & C & & D & & E
\\ \hline
$z$&$\bar{\alpha} \quad$ & $\sigma_\alpha$ & $\bar{\alpha}
 \quad $ & $\sigma_\alpha$
&$\bar{\alpha} \quad $ & $\sigma_\alpha$ &$\bar{\alpha} \quad $ 
& $\sigma_\alpha$ &$\bar{\alpha} \quad $ & $\sigma_\alpha$ 
\\ \hline
0.5& $1.00$ &  $1.01$ &  1.00 &  0.79 & $0.99$ & 1.71 & $1.00$ & 0.97
 &1.00 & 0.86\\
1&   $1.02$ &  $0.89$ &  1.00 &  0.65 & $1.03$ & 1.50 & $1.01$ & 0.86
 &1.00 & 0.65\\
2&   $0.99$ &  $0.62$ &  0.98 &  0.46 & $1.01$ & 1.01 & $0.99$ & 0.60
 & 0.98 & 0.43\\
3&   $0.99$ &  $0.51$ &  0.99 &  0.37 & $0.99$ & 0.80 & $1.00$ & 0.50
 & 0.99 & 0.34\\
4&   $0.99$ &  $0.45$ &  0.99 &  0.32 & $0.97$ & 0.68 & $0.99$ & 0.44
 & 0.99 & 0.30\\
5&   $0.99$ &  $0.40$ &  0.99 &  0.29 & $0.97$ & 0.63 & $0.99$ & 0.38
 & 0.99 & 0.27\\ 
\hline
\end{tabular}
\end{center}
\end{table}
\bigskip

\begin{table}
\caption{The average clumpiness parameter $\bar{\alpha}$ and its dispersion 
$\sigma_\alpha$ in model L with (0.2, 0.8). } 
\label{table:4}
\begin{center}
\begin{tabular}{crcrcrcrcrc} \hline \hline
 & A & & B & & C & & D & & E
\\ \hline
$z$&$\bar{\alpha} \quad$ & $\sigma_\alpha$ & $\bar{\alpha} \quad $ 
& $\sigma_\alpha$
&$\bar{\alpha} \quad $ & $\sigma_\alpha$ &$\bar{\alpha} \quad $  
& $\sigma_\alpha$ &$\bar{\alpha} \quad $ & $\sigma_\alpha$ 
\\ \hline
0.5& $1.01$ &  $1.44$ &  1.03 &  1.03 & $1.00$ &  2.49& $1.02$ &  1.35
 &0.97 & 1.08\\
1&   $1.01$ &  $1.01$ &  1.01 &  0.64 & $0.95$ &  1.78& $1.02$ &  0.96
 &0.99 & 0.60\\
2&   $1.00$ &  $0.71$ &  1.00 &  0.42 & $0.94$ &  1.23& $1.00$ &  0.68
 &0.99 & 0.36\\
3&   $1.00$ &  $0.53$ &  1.00 &  0.31 & $0.96$ &  0.92& $1.00$ &  0.51
 &0.99 & 0.26\\
4&   $1.00$ &  $0.42$ &  1.00 &  0.25 & $0.95$ &  0.73& $1.00$ &  0.41
 &0.99 & 0.21\\
5&   $1.00$ &  $0.36$ &  1.00 &  0.21 & $0.95$ &  0.63& $1.00$ &  0.35
 &0.99 & 0.17\\ 
\hline
\end{tabular}
\end{center}
\end{table}

\section{Concluding remarks}

By numerical ray-shooting in the $N$-body-simulating clumpy
cosmological models, we studied the statistical behavior of the
angular diameter distance $D_{\rm A}$ and determined the clumpiness parameter
$\alpha$ by comparing it with the Friedmann distance $(\alpha = 1)$ and 
the Dyer-Roeder distance $(\alpha > 0)$. The results show that the
average value of $\alpha$ is nearly equal to 1 and the
dispersion $(\sigma_\alpha)$ decreases with an increase of $z$, but 
that for $\theta < 1$ arcsec, $\sigma_\alpha$ is $\sim 0.5$ in model S 
and $\sim 1$ in
low-density models for $z = 1 - 2.$  Hence the influence of the 
clumpiness is not small, because the difference between distances with 
 $\alpha = 0$ and $1$ is not small for  $z > 1$. For $\theta >> 1$ 
arcsec, $\sigma_\alpha$ is so small that the distance can be regarded 
approximately as the Friedmann distance.
Therefore we can conclude for $\theta < 1$ arcsec that for rough or 
qualitative estimates of the lensing effect we can use the Friedmann
angular diameter distance $(\alpha = 1)$, but for the 
quantitative analysis of cosmological gravitational lensing, some
errors may result from using only it at the high-redshift stage ($z > 1$). 
 
For the behavior of the distribution $(N(\alpha))$, we found that
when $\sigma_\alpha < 0.5$, $N(\alpha)$ is symmetric around $\alpha =
1$, and when $\sigma_\alpha \geq 0.8$, $N(\alpha)$ is irregular, and
the symmetry is largest and smallest for S and O1 models,
respectively.
    
In the present cosmological models all particles were assumed to be
lens objects with equal masses and radii. As the number density of
particles is larger than that of visible standard galaxies, most
particles are regarded as invisible galaxies or dark matter balls. 
It is a crucial problem in cosmological lensing to specify how strong
lenses these invisible objects are.  If the particles
corresponding to only visible galaxies are regarded as lens objects, 
$\sigma_\alpha$ may be smaller than that in the above case A. If all 
visible galaxies are lens objects with $r_s = 20 h^{-1}$ kpc and 
the other particles are weaker lenses with $r_s \simeq 40 h^{-1}$ kpc, 
the resultant values for $\bar{\alpha}$ and $\sigma_\alpha$ are between
those in cases A and B.

In the above averaging process, all light rays were taken into
account. If we consider only weakly deflected light rays as
contributing to weak lensing, the dispersion  $\sigma_\alpha$ will 
be a slightly smaller than the values in the above tables. However, the
contribution of strong lensing to $\sigma_\alpha$ is small because of
its small frequency. 

In this paper we treated the case in which $m$ is on the order of the
standard galaxy. If $m$ is on the order of the rich-cluster mass, the 
influence of lensing on the distance is of course much larger and
gives much larger dispersions. However, this situation is unrealistic.   

\section*{Acknowledgements}
The author would like to thank Y. Suto for helpful discussions about
$N$-body simulations and referees for valuable suggestions.  
Numerical computations were performed on the YITP computer system.

\end{document}